\documentclass[aps,prc,preprint,showpacs,showkeys,floatfix]{revtex4}
\usepackage{graphicx}
\newcommand{\eq}[1]{Eq.~(\ref{#1})}
\newcommand{\be}{\begin{equation}}
\newcommand{\ee}{\end{equation}}
\newcommand{\bea}{\begin{eqnarray}}
\newcommand{\eea}{\end{eqnarray}}\date{\today}
\begin{document}
\preprint{NT@UW-06-???}
%%%%%%%%%%%%%%%%%%%%%%%%%%%%%%%%%%%%%%%%%%%%%%%
\title{ Pionic Color Transparency}

\author{
Arnold Larson\thanks{allars@u.washington.edu},  
Gerald A.Miller\thanks{miller@phys.washington.edu}}
\affiliation{Department of Physics, University of Washington, Seattle, WA 98195-1560}
\author{
  M. Strikman\thanks{strikman@phys.psu.edu}}
\affiliation{
Department of Physics, Pennsylvania State University, University Park, PA 16802}
%\end{tabular}

%\date{}
%\maketitle
%%%%%%%%%%

\begin{abstract}
 We use a semi-classical approximation to investigate  the effects of 
color transparency on pion electroproduction reactions.  
The resulting 
 reduced nuclear interactions   produce  significant, 
but not dominating, differences with the results of  
% $Q^2$ and $p_{\pi}$ 
%and is compared with 
conventional distorted-wave, Glauber-type treatments at kinematics
accessible to Jefferson Laboratory. Nuclear effects that could mimic the influence of
color transparency are also discussed.
\end{abstract}
\maketitle

\section{Introduction}
Color transparency is the 
reduction of initial or 
final state interactions 
in high-momentum-transfer coherent processes 
that occurs if a projectile or ejectile propagates in the nucleus as 
a  small color singlet object. See the reviews: \cite{Frankfurt:1992dx}--%\cite{Frankfurt:1994hf}
\cite{Jain:1995dd}. In such situations, the effects of emitted gluons are canceled
\cite{Low:1975sv}\cite{Nussinov:1975mw}\cite{Gunion:1976iy}
in a manner analogous to the propagation of a small electric dipole moment 
through an electrodynamic medium\cite{Bertsch:1981py} so that 
the small color singlet  behaves as a  Point Like Configuration, PLC. 
The PLC is a component of a hadron and evolves to its full size in a characteristic
hadronic time of the order of 1 fm/c. However, produced 
at sufficiently high energies, time dilation effects allow a PLC to   
propagate through the entire nucleus  before expanding. In that case, 
the production  cross sections will be larger than those computed using the standard
distorted wave or Glauber treatment.

An electroproduction  $(e,A,e'\pi X)$ experiment at 
JLAB\cite{Garrow}  attempting to measure pion transparency in nuclei has completed running 
and is currently under analysis.
 The pion electroproduction cross section is
 measured for various values of $Q^2$ and $p_{\pi}$ 
between 1 and 5 GeV and $|t| <$ 0.5 GeV$^2$.  
The experiment was done at parallel kinematics 
so that $p_{\pi}= \sqrt{Q^2+\nu^2} - k$ where $k$ is the 
magnitude of the three momentum imparted to the target.  
Targets used in the experiment include H, D, $^{12}$C, $^{64}$Cu and $^{197}$Au.  
From this data it may be possible to extract the $Q^2$, $p_{\pi}, k$ 
and A dependence of any observed color transparency.  

The state of pion transparency theory has been quiet for some time. Stimulated by
the new experiment, our  
purpose is  to display  the results of earlier theory \cite{workshop}  
using specific kinematics and nuclear targets that allows us to discuss 
the feasibility of observing color transparency in  electroproduction  reactions as well as other
nuclear effects. Papers on pionic transparency \cite{Kundu:1998ti} have appeared since 
the time of Ref.~\cite{workshop}.  The new feature of the present work is the 
evaluation at the specific experimental kinematics of \cite{Garrow}. 
We also include the effects of expansion absent from \cite{Kundu:1998ti}.  
We do not incorporate the possible interference effects between point-like and blob-like 
configurations that could produce oscillations in color transparency that are 
included in \cite{Kundu:1998ti}.

\section{The Semi-Classical Approximation}

%\vskip4cm
\begin{figure}[ht]
\label{figtc12}
\scalebox{.9}{\input{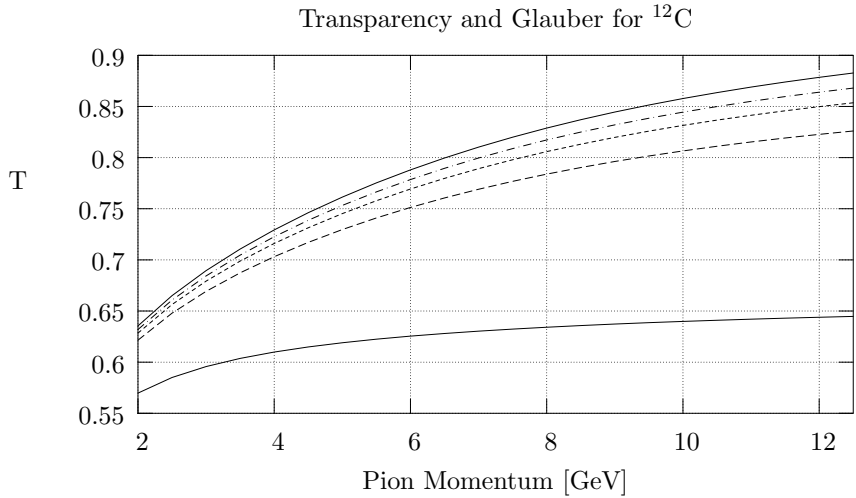}}
\caption{\small{Transparency and 
Glauber-like calculation for 
a $^{12}$C target and pion momentum up to 12 GeV.  
The top curve is for Color Transparency calculated with $\sigma_{PLC}=0$.  
The next three curves are 
for $\sigma_{PLC} \propto 1/Q^2$ evaluated at $Q^2$ = 10, 5 and 2.5 GeV$^2$ respectively.
$\Delta M^2$= 0.7 GeV$^2$. }}
\end{figure}

The final state interactions 
of semi-exclusive nuclear reactions 
can be described quite well by Glauber-type calculations.  
If a PLC is created inside the nucleus and subsequently 
evolves into a physical particle its final state interactions will be
 modified and transparency will result.

If one is summing over all nuclear final states,
the nuclear transparency can be  defined 
as the ratio of a model calculation of a nuclear cross section (including the effects of 
FSI's) to the ($A$ times the) 
cross section produced by a free nucleon target.
The use of this ratio
allows one to assess the influence of FSI's without having a detailed knowledge
of the reaction dynamics.

A semi-classical formula \cite{Farrar:1988me} has been developed to
compute nuclear 
 transparency for situations in which the kinematics of the outgoing pion are known
precisely, but the cross section involves a sum over all of the excited nuclear states
%%%ms
\cite{footnote}
%%%QUESTION - DO YOU ACTUALLY use sigma(pi N ) 25 mb or smaller?
%%%
The strength of the final state interactions depends on an emission
probability computed using the eikonal approximation and 
an effective interaction that  
parameterizes the variation of the final state interactions as the ejectile 
propagates through the nucleus.  If the particle produced
 inside the nucleus is a PLC which then expands into the observed final state, the 
interaction with the nuclear matter  deviates from that of  Glauber- type calculations.

%\vskip4cm
\begin{figure}[ht]
\label{adep}
\scalebox{.9}{\input{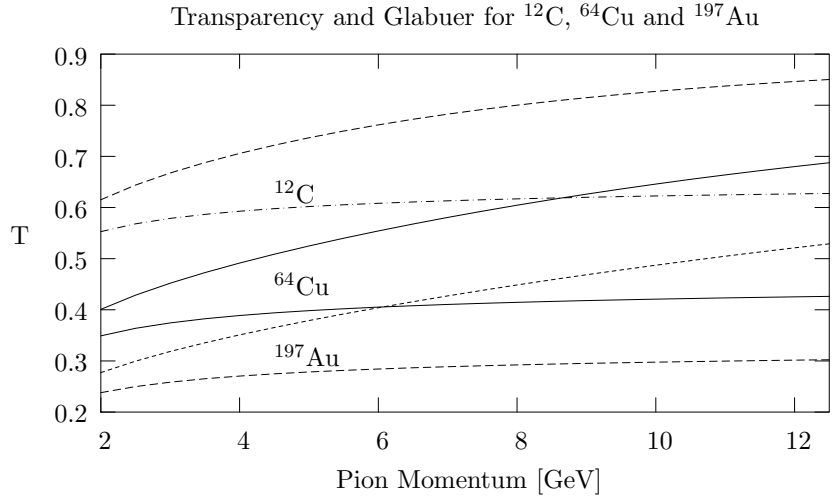}}
\caption{\small{Transparency (upper curve) and Glauber (lower curve) calculation 
for $^{12}$C, $^{64}$Cu and $^{197}$Au for pion momentum up to 12.5 GeV.  
Color Transparency is calculated with $\sigma_{PLC}$ 
evaluated at 10 GeV$^2$, and $\Delta M^2$=0.7 GeV$^2$.
Greater enhancement is to be found for larger nuclei.}}
\end{figure}

The semi-classical formula for 
 pion transparency in 
 the reaction ($e,e' A \rightarrow \pi^+ X  $)  involves
 only a single integral over the path of the outgoing pion,
\bea T= \frac{A_{eff}}{A} = \frac{1}{A} \int d^3r \rho (r) \exp[-\int_z^\infty{dz'}\sigma_{eff}(z'-z,p_\pi)
\rho( r')].\label{tsc}\eea

The nuclear density $ \rho(r)$ is of Woods-Saxon form with radius parameter 
$R=1.1 $ fm $A^{1/3}$  and diffuseness $a=0.54$ fm, and
is normalized to $A$ and the effects 
of final state interaction is contained in the effective interaction, 
$\sigma_{eff}$. 
  The effective interaction
 contains two parts, one for $z'-z$  less than a length $l_c$ 
describing the interaction 
of the expanding PLC, 
another, for larger values of $z'-z$
describing the final state
 interaction of the physical particle.  The effective interaction for the PLC is 
\bea \sigma_{eff}%^{(PLC)}
(z, p_{\pi}) = \sigma_{\pi N}(p_{\pi}) \left[\left(\frac{n^2\langle  k_t^2 \rangle}{Q^2}(1 -
 \frac{z}{l_c}) + \frac{z}{l_c} \right)\theta(l_c-z) +\theta(z-l_c)\right].
\label{sigplc}  \eea
%In this model, a PLC is created at $z_0$.  
The prediction that the interaction of the PLC will be approximately proportional to
the propagation distance $z$ for $z<l_c$ is called the quantum diffusion model.
For $z$ = 0 in \eq{sigplc} the cross section for the initially-produced PLC is identified.
% $n^2\langle  k_t^2 \rangle / Q^2$. In \eq{sigplc} the term 
\bea\sigma_{PLC}\equiv \sigma_{\pi N}(p_{\pi})\frac{n^2\langle  k_t^2 \rangle}{Q^2}\label{plcp}\eea 
% is the cross section for the initially-produced PLC with 
For the pion,  $n = 2$
and $\langle  k_t^2 \rangle^{1/2} \simeq$0.35 GeV.\cite{Farrar:1988me}   
The coherence length, $l_c$ sets the time scale for the PLC to evolve and 
determines the probability that a particle experiences reduced
 PLC interactions before leaving the nuclear matter.  For propagation 
distances $z> l_c$ 
the PLC interaction is that of a  standard final 
state approximation with $\sigma_{eff} \simeq \sigma_{\pi N}(p_{\pi})$, and is 
that of a typical  Glauber-like calculation. We take the values of  $ \sigma_{\pi N}(p_{\pi})$ from Particle Data Group  parameterization\cite{pdg}. In the limit 
 $l_c=0$ a PLC is not created 
and the calculation reduces to a Glauber-like calculation.

\vskip4cm
\begin{figure}[ht]
\label{coher}
\scalebox{.9}{\input{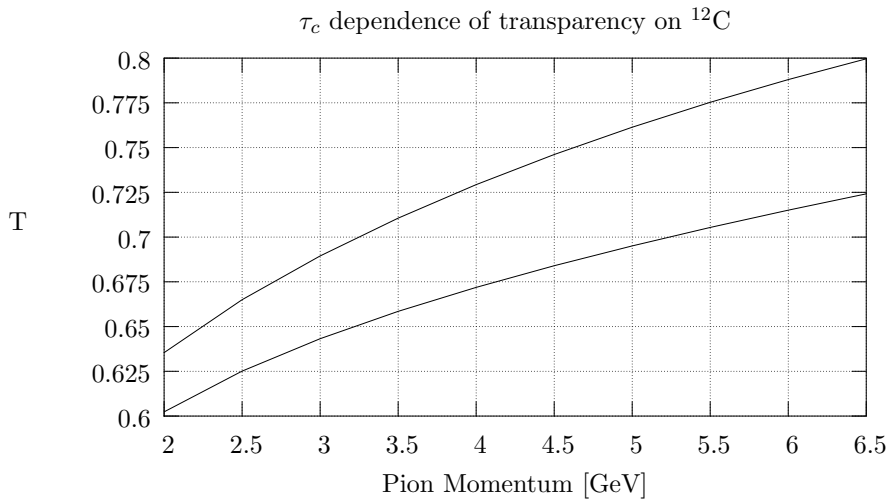}}
\caption{\small{The effects of varying the coherence length, 
$l_c = 2 p_{\pi}/ \Delta M^2$, on the transparency.  
The upper curve is obtained using  $ \Delta M^2 = 0.7$ GeV$^2$.  
The lower  curve is obtained using  $ \Delta M^2$ = 1.4 GeV$^2$.}}
\end{figure}

The transparency is known \cite{Farrar:1988me} \cite{Jennings:1990ma} to have a
strong dependence on the unknown parameter $l_c$.  The best current estimate for $l_c$
 is $2 p_h/\Delta M^2$, with $\Delta M$ given by the
 lowest lying Regge partner.  Current data do not constrain the value of 
$l_c$ for pions.   In our  calculations we use %take an optimistic value
 $\Delta M= 0.7$ GeV$^2$ which
 leads to a prediction of transparency at 
currently accessible energies, but also show results for   $\Delta M^2=1.4$ GeV$^2$.
This larger value corresponds to a very short  expansion time 
$c t_{exp}=1/\Delta M=0.16 $ fm in the rest frame, and so might be expected to provide
a reasonable estimate of a lower limit on the effects of color transparency. 
% However, currently 
%the estimate for this value is just an educated guess.

The first step is to assess the size of $T$ and its dependence on the PLC cross
section of  \eq{plcp}   for $^{12}$C.
As  shown in % Fig.~\ref{figtc12},
Fig.2,
 the effects of color transparency increase $T$ over its
value as obtained from a Glauber calculation. There is generally a small variation with $Q^2$.
The Glauber result $T\approx0.6$ 
is similar numerically to the value measured  for protons in $(e,e'p)$ with $^{12}C$, and might be
expected to be larger than that
because the $\pi$N cross section is about 2/3 the NN cross section.
However, 
the proton experiments are set up so that there is essentially only one final nuclear
state for each knocked out proton. As a result the transparency is computed using
a formula different than \eq{tsc}. Analyzing the two formulae shows that 
the chance for protons to be produced near the edge of the nucleus (where there is
little absorption) is larger than that for pions.
%%%I am not sure I agree with the discussion above - I wrote a footnote
%%where semi-classical was introduced - we can discuss on the phone.

The $A$ dependence is displayed in % Fig.~\ref{adep}.
Fig.2.
 For any fixed range of the momentum
of the outgoing pion, increasing the size of the target nucleus increases the relative 
increase in the value of $T$. 
The dependence on coherence length is shown in % Fig.~\ref{coher}.
Fig.3.
 About half (but not all)
 of the effects
of transparency are removed by decreasing the coherence length.

\begin{table}[t]
\label{kin}
\begin{center}
\caption{Kinematics of  Ref.~\cite{Garrow}.}
    \begin{tabular}{|c|c|c|c|} \hline
      $Q^2$ & $\nu$ & $k$ & $p_{\pi}$  \\ \hline\hline
      1.83  & 2.046  & 0.56 & 1.877  \\
      1     & 2.414  & 0.23 & 2.38   \\
      2.15  & 2.276  & 0.66 & 2.074   \\
      1.83  & 2.901  & 0.39 & 2.823  \\
      1.2   & 3.286  & 0.29 & 3.257  \\  
      2.15  & 3.276  & 0.46 & 3.187  \\ 
      3     & 3.48   & 0.58 & 3.315  \\
      3     & 3.576  & 0.62 & 3.418  \\
      4     & 4.096  & 0.78 & 3.857  \\
      4     & 4.126  & 0.71 & 3.897  \\
      4     & 4.235  & 0.75 & 4.016  \\
      4     & 4.732  & 0.58 & 4.571  \\
      4.8   & 4.645  & 0.88 & 4.361  \\ 
      5     & 4.822  & 0.79 & 4.54   \\ \hline
   \end{tabular}
\end{center}
\end{table}

%\vskip8cm
\begin{figure}[ht]
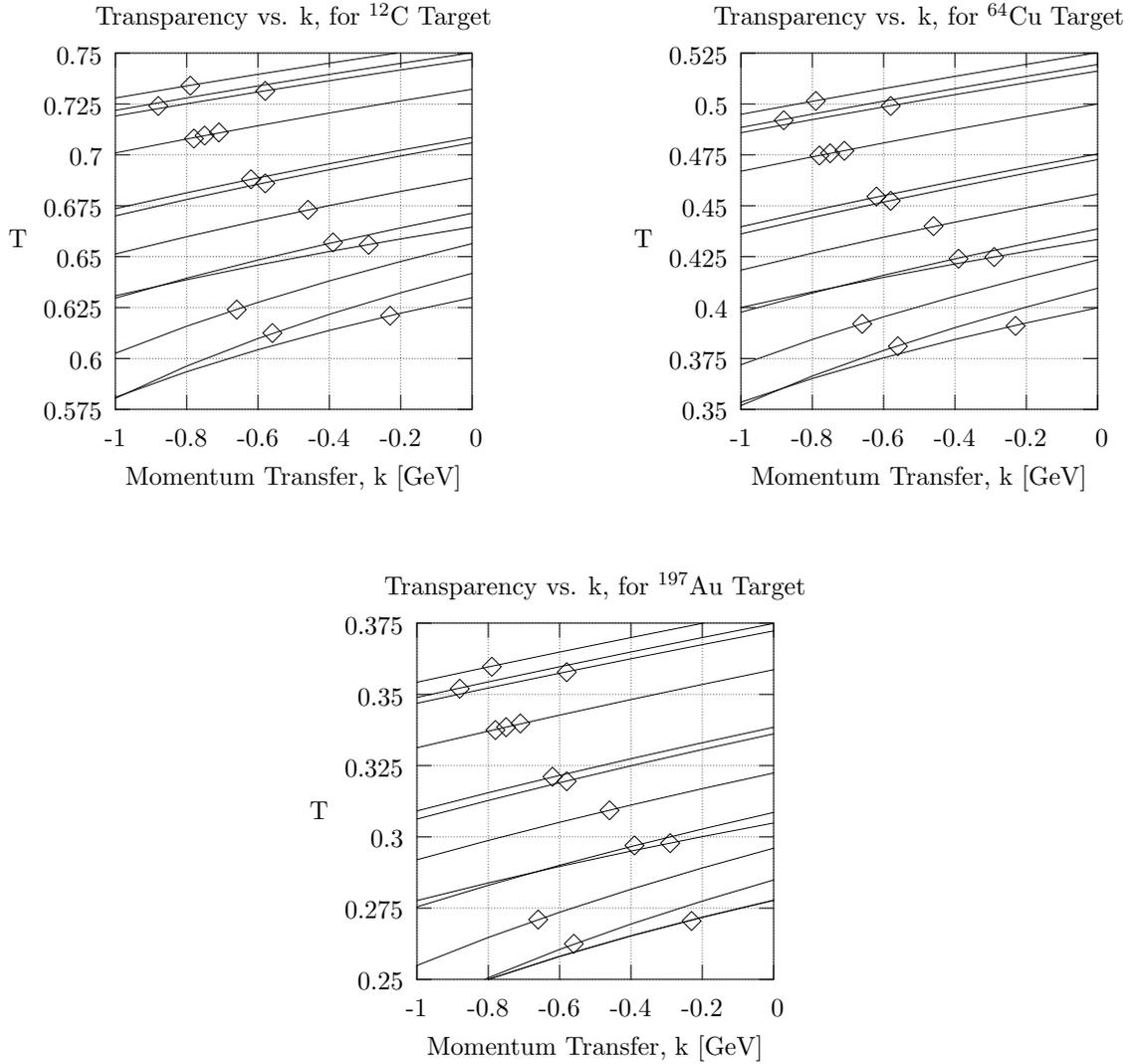

\label{expct}
\begin{tabular}{cc}
  \begin{minipage}[t]{.5\textwidth}
    \scalebox{.9}{\input{Fig4.tex}} 
    \end{minipage}
%  \hspace{0pt}
  \begin{minipage}[t]{.5\textwidth}
    \scalebox{.9}{\input{Fig5.tex}} 
    \end{minipage} \\
  \begin{minipage}[t]{0.5\textwidth}
    \par\vspace{0pt}
    \scalebox{.9}{\input{Fig6.tex}} 
    \end{minipage}
  \hspace{5pt}

  \end{tabular}
\caption{\small{Plot of Transparency vs k for a given Q$^2$ and $|\vec{q}|$ at values
 measured in the experiment at JLAB.  The curves, from bottom to top correspond 
to the values of $Q^2$ and $\nu$ listed in Table I as a function of $k$.
The diamonds show the results for the experimentally chosen values of $k$ (and therefore $p_\pi$)
  For a given $Q^2$ and $\nu$ at parallel kinematics $p_{\pi}=|\vec{q}|-k$. }}
\end{figure}

\section{ Pion Transparency at JLAB}

The experiment recently completed at JLAB measured pion electroproduction 
cross sections at specific kinematics.  The experiment used targets $^{12}$C, $^{64}$Cu and $^{197}$Au.
The essential feature of our work is to provide evaluation at the specific kinematics of
\cite{Garrow}.  The kinematics are listed in Table I % \ref{kin}. 
 and the evaluations of $T$ are shown
% in Fig.~\ref{expct}. 
in Fig.3.  
  The relevant parameters in transparency are the pion momentum, $Q^2$ and the 
size of the nucleus.  For measurements of different pion momenta and different
 nuclear targets at a given $Q^2$ one can define an enhancement factor which can 
be predicted from our semi-classical  approximation.  
It is important to emphasize here that the momenta of the pions $k$ in the kinematics of 
the Jlab experiment are large. Hence this process cannot be thought of  as a knockout of 
the pion from the meson field of the nucleon, but rather as a case of a hard process 
$\gamma^*N\to \pi N$ off a nucleon bound in the nucleus  which is governed at large $Q^2$ 
by the QCD factorization theorem \cite{Collins} which states that in this limit the process 
is dominated by the PLC configurations of the produced $q \bar q$ pair. 

%%gam
Since the pion momenta are very large compared to the characteristic scale of the nuclear momenta one 
expects nuclear modifications to the reaction to be small.  There is still a need to treat the 
modifications due to the excitation energy of the residual nucleus 
and the Fermi motion of the initial nucleon.  If the four-momentum of the struck bound nucleon
is given by ($m_N-E,{\bf p}_N$), the change in the momentum of the emitted pion
is $\approx -2.5 E-p_{N3}$, where 3 represents the photon direction. This effect shifts the peak
momentum of the produced pion by about 50 MeV/c with a spread of about 200 MeV/c.
%Two other effects are due to the Fermi motion of the struck nucleon. 
%One is the smearing of the pion spectrum for a given photon momentum. 
%%If we consider the case $p_t=0$  in the case of scattering off a free proton this 
%%corresponds just to one value of 
%%pion momenta while in the case of the scattering off the nucleon in the nucleus pion in a 
%%range of momenta can be produced.  Similarly the transverse distribution is smeared. 
The acceptance of the experiment is broad enough to 
measure pions within this momentum spread.\cite{Garrow}  
A second effect is the enhancement of the scattering 
by forward moving nucleons due to the energy dependence
of the elementary cross section. 
%%Another effect is due to the strong energy dependence of the elementary 
%%cross section - since it strongly drops with energy - 
%%the scattering by the forward moving nucleons is enhanced.
These effects are mostly important for comparing  deuteron results to those of  heavier nuclei. 
Comparing  the deuteron to heavier nuclei requires accurate  modeling of the experimental cuts.
For $A\ge 12$ the average excitation energy and average momentum become a rather weak 
function of $A$ and so one expects that the effects discussed above will mostly cancel in 
the ratio of heavy nuclei vs carbon.
 %%% clearly more should be added here based on resolution etc
  
In the following $T$ is plotted vs. $k$, the magnitude of the three-momentum exchanged 
with the target. 
For parallel kinematics, the magnitude of the pion three-momentum 
is $p_{\pi}$ = $|\vec{q}| - k$.
The usual color transparency prediction is for greater transparency with greater 
final state momentum.  
Thus the chosen value of $k$ strongly influences the computed values of $T$.
For the experimental data one can look at the greatest ratio of 
transparencies for the collected data and look for any variation in this ratio
 between the different targets.

There seems to be enough variation with $Q^2$ and $A$ to allow identification
of color transparency for the current beam energies. However, the use of a higher 
energy beam planned
for JLab12 would greatly extend the scope of the study as well as the possibility 
of obtaining a clean
observation of  the effects of color transparency.

%%gam

There are other significant effects that depend on the value of $k$  that are not included in the present calculation. The nuclear EMC  effect
tells us that nuclear interactions act to suppress the effects of  quarks moving with large Bjorken $x$. Therefore the chance for a pion
of large momentum, $k$, to exist in a nucleus could depend on the  nuclear density and differ strongly between the deuteron
and heavier nuclei. However, the EMC effect is similar for all  nuclei with mass as great as Carbon. Therefore comparisons between
heavier nuclei could be more useful in extracting information about  color transparency than comparing between heavy nuclei and
the deuteron.

\section{Conclusion}

Figure 4 shows a clear pattern in the $Q^2$,  $A$  and $k$ (three-momentum 
transfer to the nucleus) dependence of the transparency $T$.
Use of the $A$-dependence could allow the identification of color transparency at the current
energies of JLab. The use of a higher energy beam expected at JLab12 would 
significantly enhance the probability of observe
 color transparency in pion electroproduction experiments. If the dependence of the 
cross section on $k$ does not vary greatly between different nuclei than an 
enhancement of the cross section for a given $k$
 will be a clear signal of color transparency. We may expect that this variation will be 
similar
for heavy nuclei but not for light nuclei. The ratios of cross sections for heavy nuclei can
therefore provide more reliable information than considering ratios with the deuteron 
cross section.
Any understanding of transparency can not be complete without an understanding of the $k$ 
dependence of the $(e,e'\pi)$ reaction. This will be pursued elsewhere.

%\bibliography{bibtemplate.bib}

\begin{thebibliography}{99}
\bibitem{footnote}
A quantum mechanical Glauber type procedure is to calculate absorption 
on the amplitude level and then
 to sum over the final states of the 
nuclear system, see e.g. \cite{zhalov89}. This lengthy procedure %which is rather
%lengthy 
gives a somewhat smaller value of the transparency in the Glauber limit. 
However the rate of the change of the 
transparency due to inclusion of 
the color transparency effects is rather similar. Thus 
we will use  the semi-classical formula here for simplicity. 
Note also that in the quantum mechanical formula $\sigma$ 
reaches the value of $\sigma_{tot}$ while in the 
semi-classical formula it reaches $\sigma_{inel}$.
\end{thebibliography}

%\vskip6cm
\bibliographystyle{unsrt}

\end{document}